\documentclass[reprint,superscriptaddress,amsmath,amssymb,aps,prb,longbibliography,nofootinbib
]{revtex4-2}

\usepackage{graphicx,float}% Include figure files
\usepackage{dcolumn}% Align table columns on decimal point
\usepackage{bm,bbm,mathrsfs}% bold math
\usepackage{soul}%for editing strike out
\usepackage{xcolor}
\usepackage[colorlinks=true,linkcolor=blue,citecolor=blue]{hyperref}

\newcommand{\bra}[1]{\left\langle#1\right|}
\newcommand{\ket}[1]{\left|#1\right\rangle}

\begin{document}

\preprint{APS/123-QED}

\title{Delicate Topology of Luttinger Semimetal}

\author{Penghao Zhu} 
\affiliation{Department of Physics and Institute for Condensed Matter Theory, University of Illinois at Urbana-Champaign, Urbana, Illinois 61801, USA}
\author{Rui-Xing Zhang}
\affiliation{Department of Physics and Astronomy, The University of Tennessee, Knoxville, TN 37996, USA}
\affiliation{Department of Materials Science and Engineering, The University of Tennessee, Knoxville, TN 37996, USA}
\affiliation{Institute of Advanced Materials and Manufacturing, The University of Tennessee, Knoxville, TN 37920, USA}

\date{\today}

\begin{abstract}
Recent advances in delicate topology have expanded the classification of topological bands, but its presence in solid-state materials remains elusive. Here we show that delicate topology naturally emerges in the Luttinger-Kohn model that describes many semiconductors and semimetals. In particular, the Luttinger semimetal is found to be a quantum critical point leading to a quantized jump of an integer-valued delicate topological invariant. Away from this criticality, we have identified new types of electronic insulators and semimetals with intertwined stable and delicate topologies. They all carry gapless surface states that transform anomalously under rotation symmetry. Our work provides a starting point for exploring delicate topological phenomena in quantum materials.
\end{abstract}
\maketitle

%\tableofcontents

\section{Introduction}
Topological band insulators are not smoothly connected to trivial, atomic ones~\cite{Hansan:2010,qi2011topological,bansil2016colloquium}. To delineate this topological difference, a well-accepted measure is Wannierizability - the ability to describe the occupied subspace of a system using symmetric and localized Wannier functions ~\cite{Thonhauser2006Chern,soluyanov2011Z2Wannier}. Atomic insulators are inherently Wannier representable, setting them apart from all previously known topological insulators (TIs) that are not. This Wannier obstruction in TIs can be further classified into two types: stable and fragile. Stable obstructions persist even when additional trivial bands are added to the valence-band manifold, while fragile obstructions can be nullified by such an addition~\cite{po:2018fragile}. Over the past decade, this paradigm of Wannierizability has been a cornerstone in the modern frameworks of topological bands~\cite{bradlyn2017topological,slager2017bands,po2017symmetry}, such as symmetry indicator theory and topological quantum chemistry, which subsequently leads to a triumph in identifying tens of thousands of candidate topological materials~\cite{vergniory:2019complete,zhang2019catalogue,tang2019comprehensive}.

Are Wannierizable systems always trivial? Recently, the answer to this fundamental question is revised by the discovery of {\it delicate topology} in Wannierizable systems~\cite{nelson:2021multi}. Specifically, delicate TIs possess ground-state wavefunctions that are exponentially localized but cannot be completely confined within a single primitive unit cell. Notably, crystalline symmetries can help mandate this ``multicellularity" of Wannier functions in real space by enforcing an integer-quantized polarization difference between a pair of high symmetry lines in the $k$-space Brillouin zone (BZ). This phenomenon is dubbed a {\it returning Thouless pump} (RTP), which physically enables exotic consequences such as gapless surface states with an angular-momentum anomaly~\cite{nelson:2021delicate,zhu:2023spinhopf}, a large bulk photovoltaic effect~\cite{alex:2022topological}, etc. Despite the rapidly growing conceptual progress \cite{zhu:2023scattering,lim2023real,chen:2023chern,tyner:2022dipolar,brouwer:2023homotopic,graf:2022multifold}, however, delicate topological bands are invisible to conventional band diagnoses such as symmetry indicators and band representations. This significantly hinders us from tracing delicate topological physics in solid-state materials.

Our main finding is that the classic Luttinger-Kohn (LK) Hamiltonian~\cite{Luttinger1955} precisely serves as a {\it a critical point for delicate topology}. As a textbook example for ${\bf k \cdot p}$ models, the three-dimensional (3D) LK Hamiltonian captures the 4-fold band degeneracy at the BZ origin $\Gamma$ in various semiconductors (e.g., Si and Ge) and semimetals (e.g., $\alpha$-Sn~\cite{groves1963Tin} and HgTe~\cite{bernevig2006quantum}). We will primarily focus on the semimetallic scenario, i.e., a Luttinger semimetal (LSM)~\cite{moon2013luttinger,kondo2015quadratic,hu2023topological}, where this degeneracy appears as a quadratic band touching (QBT) of $j=3/2$ fermions. Since the early stage of TI research, LSMs have been extensively studied due to their proximities to $\mathbb{Z}_2$ TIs~\cite{dai2008HgTe}, Dirac~\cite{xu2017Sn} and Weyl semimetals~\cite{cano2017weyl}. However, unlike Dirac and Weyl fermions, the QBT does not yield an intuitive topological-charge interpretation. Therefore, it remains unclear whether the LSM itself would inherently encode any topological information.   

The unexpected connection between delicate topology and LSM originates from the following observation: a smooth evolution path ${\cal P}$ can be established between the LK Hamiltonian and a Dirac dipole (DD) Hamiltonian, wherein any node-enclosing Fermi surface remains gapped. The DD has recently been recognized as a delicate topological critical point (TCP) for spin Hopf systems~\cite{zhu:2023spinhopf}, leading us to anticipate similar behavior in the LSM. To validate the topological equivalence between LSM and DD, we have quantitatively confirmed the existence of a new topological charge $w_{\text{net}}\in\mathbb{Z}$. Remarkably, this charge remains both quantized and conserved along the entire ${\cal P}$ path, which provides a natural topological interpretation for the LSM. We further perturb the LK Hamiltonian in the continuum limit and two adjacent phases (dubbed A and B) emerge in the topological phase diagram. Notably, any path directly connecting A and B would pass the LSM and necessarily involves a quantized jump of the returning Thouless pump invariant, as justified by our Wilson-loop calculations. This directly proves LSM to be a delicate TCP. 

The topological nature of phases A and B are explored by converting the LK Hamiltonian into a lattice model. By employing various lattice regularization schemes, we have successfully identified the anticipated delicate TI phase, as well as intriguing scenarios where delicate and stable topologies coexist and interact synergistically. As proof of concept, we discuss two examples based on the LK Hamiltonian, a weak TI and a Weyl semimetal (WSM). In both cases, stable topological invariants, such as $\mathbb{Z}_2$ indices and Weyl charges, decide the number and $k$-space locations of the boundary states. Meanwhile, the delicate topology requires the top and bottom surface states to carry different angular momentum labels $J_z$, resulting in a remarkable surface angular-momentum anomaly. Across the delicate topological transition through the QBT, the $J_z$ pattern of surface states will vary as a response to the change in RTP, while the stable topological information remains invariant. We term this novel class of topological phenomena {\it RTP-enriched topological phases.}
        
The remainder of the article is organized as follows. In Sec \ref{sec:topoFS}, we define and discuss the topological charge carried by the QBT of LK Hamiltonian. To facilitate our subsequent discussions, a brief review of RTP physics is provided in Sec \ref{sec:critical}. We then establish the QBT of LK Hamiltonian as a delicate TCP and further discuss the phase diagram surrounding this TCP. A representative delicate TI phase is identified and analyzed in Sec.~\ref{sec:dti}, elaborating on the origin of its unconventional surface states. We further introduce the concept of RTP-enriched topological phases, which is exemplified through investigations on the weak TI and WSM phases. We conclude with highlights, remarks on candidate materials, and open questions for future research in Sec \ref{sec:conclusion}.

%%%%%%%%%%%%%%%%%%%%%%%%%%%%%%%%%%%%%%%%%%%%%%%%%%%%%%%%%%%%%%%%%%%%%%%%%%

\section{Topological Charge of LSM}
\label{sec:topoFS}

\begin{figure*}[t]
\centering
\includegraphics[width=0.95\textwidth]{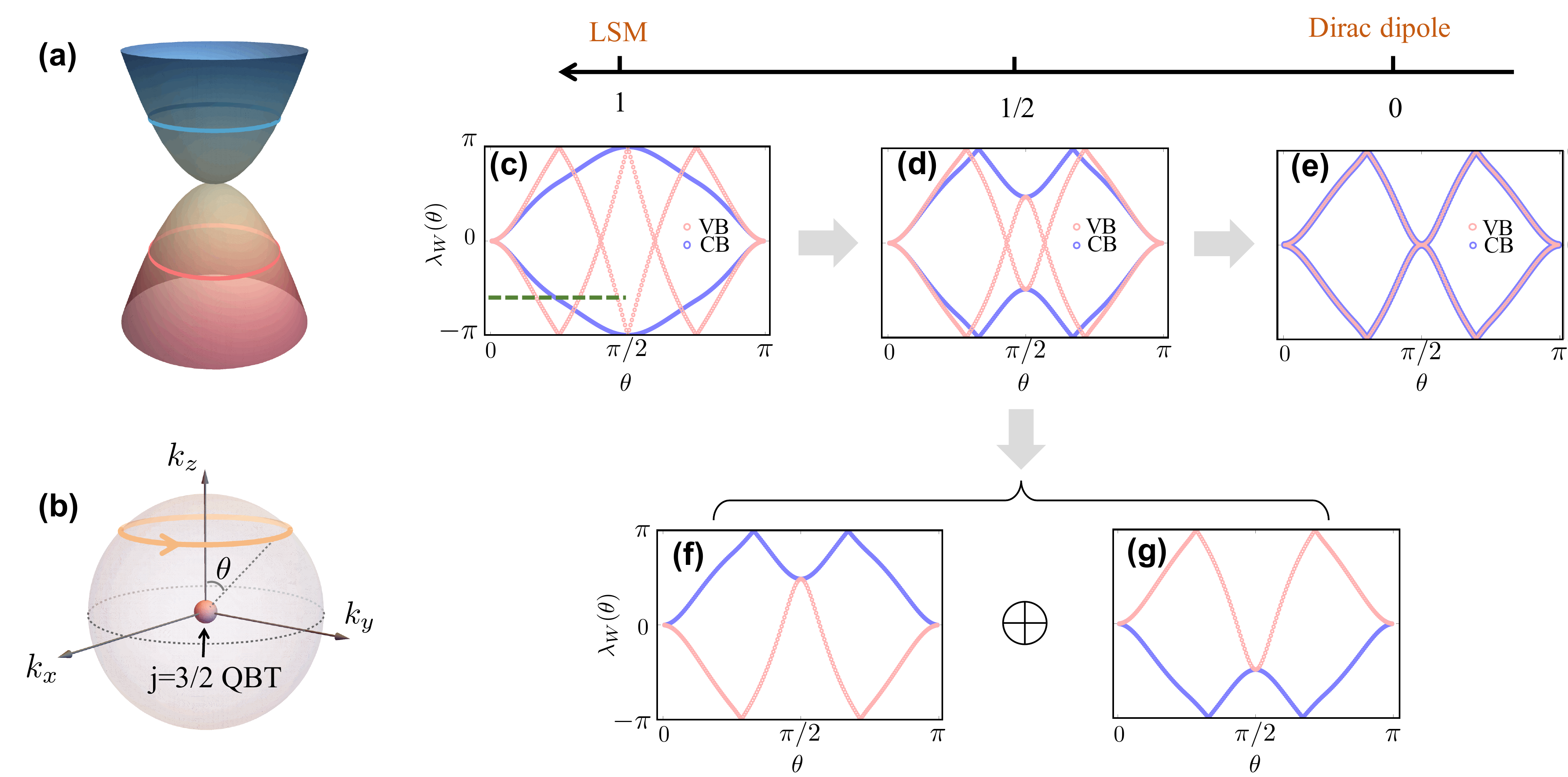}
\caption{(a) the QBT of a LSM. The blue and red rings schematically denote conduction and valence Fermi surfaces for a fixed $|{\bf k}|$, based on which we will calculate the Berry phases. (b) shows a $k$-space Fermi sphere surrounding the $j=3/2$ QBT, with $\theta$ labeling the latitude. We follow the orange loop to calculate the 1D Wilson loop spectrum for every $\theta$. The results for both conduction (blue circles) and valence (red circles) electrons are shown for (c) LSM, (d) an intermediate $h_{0}^{(\alpha=1/2)}$, and (e) Dirac-dipole. Following the green dashed reference line in (c) and the Wannier-band decomposition scheme depicted in (f) and (g), it is clear that all three cases feature a net winding number $w_\text{net}=4$ that is contributed by both CB and VB subspaces.}
\label{fig:FSwinding}
\end{figure*}

The QBT of $j=3/2$ fermions in LSM systems arises from a 4D double-valued irreducible representation (irrep) of crystalline point groups such as $O$, $T_d$, and $O_h$. The low-energy physics is well captured by the 4-band LK Hamiltonian,
\begin{equation}
\label{eq: HLSM}
H_{LK}({\bf k})=\left(\lambda_1+\frac{5}{2} \lambda_2\right) \mathbf{k}^2-2 \lambda_2(\mathbf{k} \cdot \mathbf{S})^2,
\end{equation}
under the $|j_z\rangle$ basis $\Psi = (|\frac{3}{2}\rangle, |\frac{1}{2}\rangle, |-\frac{1}{2}\rangle, |-\frac{3}{2}\rangle)^T$. Here ${\bf S} = (S_x, S_y, S_z)$ denote the spin-3/2 Pauli matrices. Besides respecting time-reversal symmetry $\mathcal{T}=i\sigma_{x}s_{y}K$, $H_{LK}$ preserves an $O(3)$ symmetry in the continuum limit, which results in an isotropic dispersion relation with $E = (\lambda_1 \pm 2 \lambda_2) |{\bf k}|^2$. When $|\lambda_1|<2|\lambda_2|$, $H_{LK}$ describes the quadratically dispersing LSM phase. Without loss of generality, we will focus on this LSM phase and set $\lambda_1=0$ and $\lambda_2=1$ by default, unless specified otherwise.   

Let us recall that any $k$-space Fermi surface containing a Weyl node will harbor a Berry flux of $\pm 2\pi n$ with $n\in\mathbb{Z}$~\cite{armitage2018rmp}, which is further translated to a quantized monopole charge following Gauss's Law. Inspired by the same spirit, we consider a similar spherical Gaussian surface ${\cal S}$ that encloses the QBT of LSM in $k$ space, as illustrated in Fig.~\ref{fig:FSwinding}(b). If the QBT indeed carries some topological charge ${\cal Q}$, we expect ${\cal Q}$ to induce an unconventional pattern of the Berry phase/curvature on ${\cal S}$.
To verify this, we denote a parallel of latitude on the Gaussian sphere as ${\cal L}_\theta$ [orange line in Fig.~\ref{fig:FSwinding}(b)], where $\theta\in [0,\frac{\pi}{2}]$. We further define a Wilson loop operator $\hat{W}(\theta)=\prod_{\mathbf{k}\in {\cal L}_\theta} {\cal P}_{\mathbf{k}}$ along every ${\cal L}_\theta$~\cite{king-smith1993polarization}. Here, ${\cal P}_{\mathbf{k}}=\sum_{n\in v/c}\ket{u_{n}(\mathbf{k})} \bra{u_{n}(\mathbf{k})}$ is the projector onto valence/conduction subspace. The Wannier band spectrum $\lambda_{W}(\theta)$ is the set of exponents of unit modulus complex eigenvalues of the Wilson loop operator, which intuitively describes the evolution of 1D Berry phase along ${\cal L}_\theta$ for the projected subspace.   

In Fig.~\ref{fig:FSwinding}(c), we plot the Wannier bands evolving from the north pole ($\theta=0$) to the south pole ($\theta=\pi$). It is worth emphasizing two facts here: (i) $(\lambda_W, \theta) \in (-\pi,\pi)\times [0,\pi]$ span a tube geometry; and (ii) there exist two Wannier bands in both conduction and valence space. The pattern of Wannier bands is quantified by a {\it relative winding number} $w$, which is defined by the number of intersections between the Wannier bands and a horizontal reference line from $0<\theta<\pi/2$ sitting at an arbitrary value of $\lambda_W$ [e.g., the green dashed line in Fig.~\ref{fig:FSwinding} (c)]. Our calculation in Fig.~\ref{fig:FSwinding} (c) clearly unveils nontrivial relative windings with $w_{c}=1$ for conduction bands (CBs) and $w_{v}=3$ for valence bands (VBs), respectively.      

To better decipher the topological physics here, we further propose a generalized LK model,
\begin{equation}
\label{eq:generalizedLSM}
\begin{aligned}
    h_0^{(\alpha)}({\bf k}) &= 
    -\sqrt{3} \alpha (k_x^2-k_y^2) \gamma_1 - 2\sqrt{3} \alpha k_x k_y \gamma_2 
    \\
    &-2\sqrt{3} k_z k_x \gamma_3 - 2\sqrt{3} k_z k_y\gamma_4 + M({\bf k}) \gamma_5,
\end{aligned}
\end{equation}
where $h_0^{(1)} = H_{LK}$ recovers the original LK Hamiltonian. Here, $\gamma_1 = \sigma_x \otimes s_0$,  $\gamma_2 = \sigma_y \otimes s_0$, $\gamma_3 = \sigma_z \otimes s_x$,
$\gamma_4 = \sigma_z \otimes s_y$, $\gamma_5 = \sigma_z \otimes s_z$,
and $M({\bf k}) = k_x^2 + k_y^2 - 2k_z^2$. 
$\sigma_{x,y,z}$ and  $s_{x,y,z}$ are Pauli matrices for orbitals and spins of electrons, respectively. We consider an evolution path from the LSM limit $\alpha=1$ to $\alpha=0$, along which we find that (i) the QBT of $j=3/2$ fermions remains resilient; (ii) an energy gap is always present for any finite-radius Fermi sphere ${\cal S}$ that encloses the node. We, therefore, expect that any topological information of the QBT should be preserved as $\alpha$ evolves. 

Interestingly, $\alpha=0$ represents another special limit with $w_c = w_v =2$, as shown in Fig.~\ref{fig:FSwinding} (e). The quantized Wannier-band windings for $h_0^{(0)}$ can be interpreted as a {\it Dirac dipole} (DD), where both the upper and lower hemispheres of the Fermi sphere ${\cal S}$ individually harbor a quantized Kane-Mele $\mathbbm{Z}_{2}$ charge~\cite{kane2005Z2}. Indeed, $h_0^{(0)}$ formally reproduces the DD Hamiltonian constructed in Ref.~\cite{zhu:2023spinhopf}, up to some parameter renormalization. 

Away from $\alpha=0$ or $1$, however, the Wannier bands $\lambda_{W}(\theta)$  in both valence and conduction space do not generally admit an integer number of windings. This is clearly shown in the Wannier band spectrum for $\alpha=1/2$ in Fig.~\ref{fig:FSwinding} (d). Importantly, despite the distinct patterns in Figs.~\ref{fig:FSwinding} (c) - (e), a general $h_0^{(\alpha)}$ always carries an intriguing Wannier-band degeneracy between CB and VB electrons at the equator of ${\cal S}$ ($\theta=\pi/2$). Since the equator respects a mirror symmetry $M_z:z\rightarrow -z$, the corresponding Wannier bands can be labeled as $\lambda_{W,c/v}^{\eta}(\alpha)$, where $\eta=\pm$ labels different mirror sectors. The Wannier-band degeneracy is described by     
\begin{equation}
\label{eq:lambdaequator}
\begin{aligned}
\lambda_{W,v}^{+}(\alpha) &=\lambda_{W,c}^{-}(\alpha)
=-\lambda_{W,v}^{-}(\alpha)=-\lambda_{W,c}^{+}(\alpha)
\\
&=2\pi\left(1-\frac{1}{\sqrt{1+3\alpha^2}}\right) \ \ \ \text{mod}\ 2\pi.
\end{aligned}
\end{equation}
In the Supplemental Material (SM) \ref{app:wlosymmetry}, we derive Eq.~\eqref{eq:lambdaequator} analytically and further prove that the above degeneracies are stabilized by a combination of time-reversal, spatial inversion, and $M_z$ symmetries. Besides, there also exist multiple robust crossings of Wannier bands at $\lambda_W(\theta) = \pm \pi$ that are protected by the rotation symmetry along $z$-direction~\cite{zhu:2023spinhopf}.

Inspired by the above observations, we pursue a new strategy to topologically describe the QBT of $j=3/2$ fermions by treating the CB and VB Wannier bands as a whole. Following Fig.~\ref{fig:FSwinding} (f) and (g), we can pair every CB Wannier band with a VB one and such a group of Wannier bands always shows a double relative winding. Bringing two groups together, we find a quantized relative winding number contributed by both CB and VB subspaces with
\begin{equation}
    w_\text{net} = 4.
\end{equation}
Crucially, Eq.~\eqref{eq:lambdaequator} dictates that an increase of winding for CB electrons will necessarily lead to the same amount of winding reduction for the VB space, so that $w_\text{net}$ remains both {\it quantized} and {\it conserved}, irrespective of $\alpha$. This strictly guarantees that the Wannier bands for CB and VB electrons on ${\cal S}$ cannot be unwinded simultaneously. Physically, $w_\text{net} \neq 0$ implies an intrinsic topological obstruction that prevents ${\cal S}$ from being deformed into a trivial Berry-phase-free Fermi surface.

\section{LSM as a Delicate Topological Critical Point}
\label{sec:critical}

The above results have clearly unveiled an inherent connection between LSM and delicate topology. First of all, $w_\text{net}$ is defined for the entire Hilbert space, which contrasts with most previously known topological indices that are solely based on VB properties. The quantization of $w_\text{net}$ could then be spoiled when coupling an additional set of trivial bands to either CB or VB subspace, resembling the definition of delicate topology~\cite{nelson:2021multi}. This possibility is numerically verified in the SM~\ref{app:delicacy} for the LK Hamiltonian. Hence, $w_\text{net}$ should be interpreted as a {\it delicate topological charge} for the QBT of $j=3/2$ fermions. 

Furthermore, the conservation of $w_\text{net}$ indicates a topological equivalence between the LSM ($\alpha=1$) and the DD ($\alpha=0$). Notably, the DD band degeneracy is found to be a phase transition point between two class-AII crystalline insulators with distinct delicate topologies~\cite{zhu:2023spinhopf}. It is then natural to expect the LSM to carry similar physics.

\begin{figure*}[t]
\centering
\includegraphics[width=2\columnwidth]{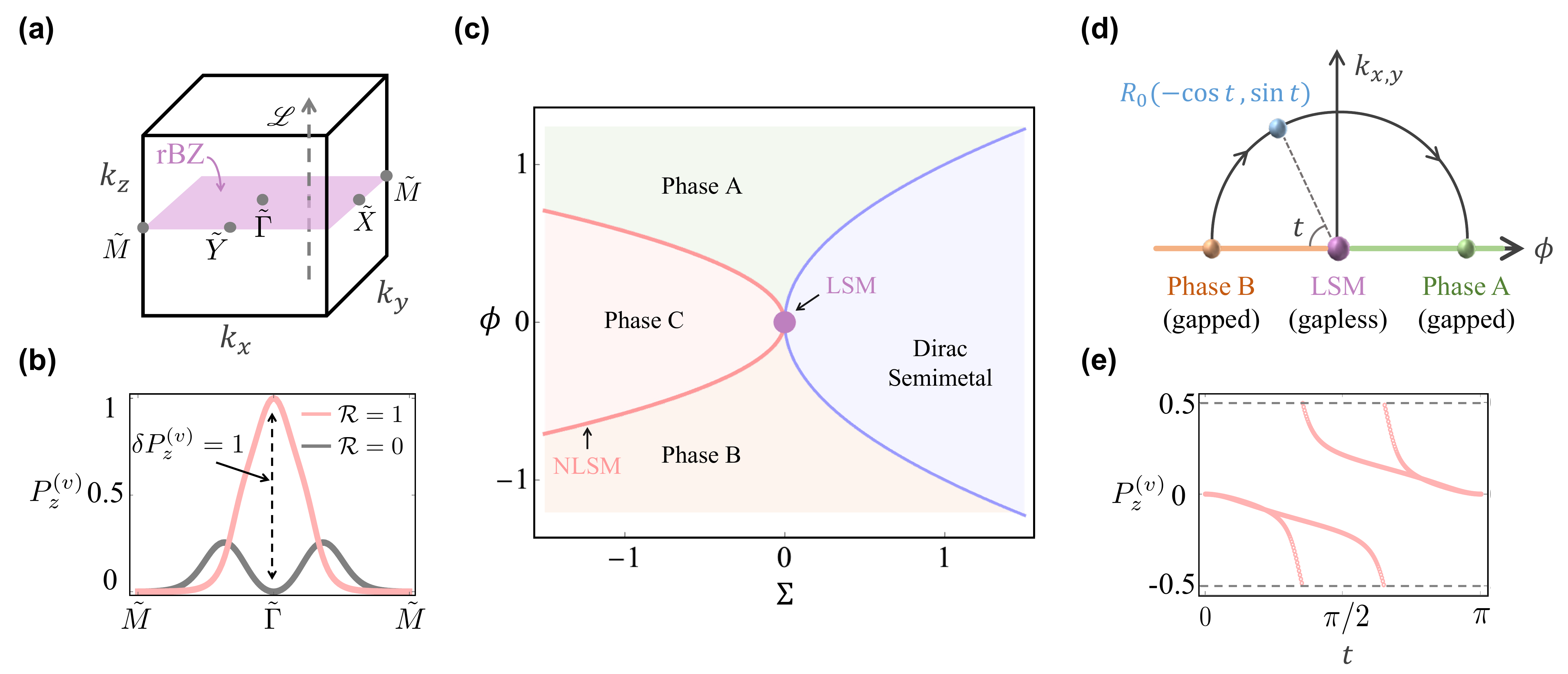}
\caption{(a) illustrates the rBZ. $\tilde{M}$ and $\tilde{\Gamma}$ are rotation-invariant momenta. $\mathscr{L}$ shows a $k_z$ loop along which we will calculate $P^{(v/c)}_{z}$. (b) shows the polarization evolution for both a delicate TI (${\cal R}=1$) and a trivial insulator (${\cal R}=0$), the transition between which will involve a quantized jump of $P_{z}$ at $\tilde{\Gamma}$. The topological phase diagram of a perturbed LSM is shown in (c) as a function of $\Sigma$ and $\phi$. (d) describes a smooth path $\mathscr{R}$ in the parameter space $(\mathbf{k},\phi)$, along which we track the $P_{z}^{(v)}$. (e) clearly reveals a quantized change of $P^{(v)}_{z}$ from phase B to phase A, further establishing the LSM as a delicate TCP.}
\label{fig:phasediagram}
\end{figure*}

\subsection{Theory of Returning Thouless Pump}
\label{sec:reviewRTP}
As a warm-up, let us first review the theory of RTP, which is a delicate topological phenomenon that will play a key role in deciphering the core physics of LSM. Without loss of generality, we consider a 3D {\it Wannierizable} $C_4$-symmetric system with the rotational axis aligning with the $z$-axis. A theory for general $C_n$ symmetry can be defined analogously. To reveal the hint of delicate topology, we consider the 1D charge polarization $P^{(\kappa)}_{z}({\bf k}_{\parallel})$ for each ${\bf k}_\parallel$-labeled $k_z$ axis, where the index $\kappa=v/c$ labels the valence/conduction electrons. The 2D $k_{x}-k_{y}$ plane that ${\bf k}_\parallel$ lives in is dubbed a reduced Brillouin zone (rBZ) [see Fig.~\ref{fig:phasediagram}(a)]. In the Berry-phase formalism, we have
\begin{equation}
\label{eq:polarization}
P^{(\kappa)}_{z}({\bf k}_\parallel)=\oint\frac{dk_{z}}{2\pi} \operatorname{Tr}[\mathcal{A}_z^{(\kappa)}({\bf k}_\parallel,k_{z})],
\end{equation}
where $(\mathcal{A}_z^{(\kappa)})_{mn}=i\langle u^{(\kappa)}_{m}|\partial_{k_{z}}|u^{(\kappa)}_{n}\rangle$ is the non-Abelian Berry connection along $z$-direction, and $|u^{(\kappa)}_{n}\rangle$ is the $n$-th valence or conduction Bloch states.

There exist two $C_4$-invariant axes with $\tilde{K}_0 \in\{\tilde{\Gamma},\tilde{M}\}$, along which electrons are further labeled by an angular momentum index $J_z$ modulo $4$. It is then possible for the system to respect a mutually disjoint condition (MDC), that for any given $J_z$, all of $J_z$-labeled eigenstates at a $\tilde{K}_0$ must live in either conduction or valence subspace, but not in both. The MDC guarantees that the following polarization difference is quantized in the unit of lattice constant~\cite{nelson:2021multi,nelson:2021delicate}, with
\begin{equation}
\begin{aligned}
    &{\cal R} \equiv P^{(v)}_{z}(\tilde{\Gamma}) - P^{(v)}_{z}(\tilde{M}) 
    \\
    &=-(P^{(c)}_{z}(\tilde{\Gamma}) - P^{(c)}_{z}(\tilde{M})) \in \mathbb{Z}.
    \label{eq:RTP_def}
\end{aligned}
\end{equation}
Viewing $\tilde{M}-\tilde{\Gamma}-\tilde{M}$ as an adiabatic evolution path for $P^{(v)}_z$, we schematically depict the scenario for both ${\cal R}=0$ and ${\cal R}=1$ in Fig.~\ref{fig:phasediagram} (b). We note that: (i) from $\tilde{M}$ to $\tilde{\Gamma}$, there will be ${\cal R}$ electric charges pumping from one unit cell to its neighbor cell along $z$ direction, just like a regular Thouless pump; (ii) the Thouless pump process will be exactly reversed for the second half of the cycle, i.e., from $\tilde{\Gamma}$ to $\tilde{M}$. Therefore, ${\cal R}$ physically describes a returning Thouless pump (RTP) over the complete evolution cycle.  

From the real-space perspective, a non-zero ${\cal R}$ directly implies that the ground-state wavefunctions necessarily span at least $|{\cal R}|$ primitive unit cells. This strictly prevents a RTP insulator from being smoothly deformed into an atomic insulator. While the value of RTP invariant ${\cal R}$ is robust against any adiabatic deformation that respects $C_4$ and the MDC, it is always possible to spoil it through an MDC violation, by coupling either the valence or conduction bands with a certain set of atomic bands. In this sense, the topology described by an RTP is ``delicate". 

Consider a transition of RTP tuned by a single parameter $\phi\in\mathbb{R}$, with ${\cal R}(\phi>0) \neq {\cal R}(\phi<0)$. Following Eq.~\ref{eq:RTP_def}, it is easy to see that a RTP jump $\Delta {\cal R}$ across the TCP is induced by a quantized jump of charge polarization at $\tilde{\Gamma}$, $\tilde{M}$, or both. Specifically, 
\begin{equation}
    \Delta {\cal R} = \Delta P^{(v)}_{z}(\tilde{\Gamma}) - \Delta P^{(v)}_{z}(\tilde{M}),
\end{equation}
where $\Delta P^{(v)}_{z}(\tilde{K}_0) \equiv P^{(v)}_{z}(\tilde{K}_0) |_{\phi=0^+} - P^{(v)}_{z}(\tilde{K}_0)|_{\phi=0^-}$. In the forthcoming subsection, we will exploit the quantized polarization jump as a key indicator for delicate topological phase transitions.

\subsection{Quantized Polarization Jump through the QBT in LSM}

To uncover the hidden RTP physics of LSM, we consider perturbing $H_\text{LK}$ with,
\begin{equation}
\label{eq:perturbphi}
 H_\phi ({\bf k}) = -2\sqrt{3}\phi (k_x \gamma_{35} + k_y \gamma_{45}) + (\Sigma - \beta \phi^2) \gamma_5,
\end{equation}
with $\gamma_{jk}=-i\gamma_j\gamma_k$ and $\beta \geq 0$. The first term of $H_\phi(\mathbf{k})$ manifests as a 2D Rashba effect, and the second term describes an in-plane lattice-strain effect. While respecting time-reversal symmetry and the continuous rotation symmetry $C_{\infty}$ around $z$ axis, $H_\phi$ breaks spatial inversion, $M_z$, and all $n$-fold rotation symmetries around $x$ or $y$ axis with $n>2$. The phase diagram of $H_c \equiv H_\text{LK}+H_\phi$ for a fixed $\beta=1$ is shown in Fig.~\ref{fig:phasediagram} (c). We analytically find three insulating regions, which we name phases A, B, and C.  These insulating phases are separated by two gapless regions: (i) A Dirac semimetal phase occurs when $\Sigma\geqslant \beta \phi^2$; and (ii) a nodal-loop semimetal (NLSM) state forms at $\Sigma = (\beta - 4)\phi^2$, which has a 2-fold-degenerate nodal loop at
$k_x^2+k_y^2 = 4\phi^2$ and $k_z=0$. A detailed derivation of the above gapless conditions can be found in the SM \ref{app:gapless}.

Notably, to connect phases A and B without passing any intermediate phases, any evolution path will {\it necessarily} hit the LSM degeneracy at $\Sigma=\phi=0$.  The equivalence between LSM and DD inspires us to check whether phases A and B could differ by a $\Delta {\cal R}$. First, since $H_c({\bf k})$ is an effective model around $\Gamma$, we will focus on the contribution of $\Delta {\cal R}$ only from $\Gamma$, but not other high-symmetry momenta. Namely, the continuum limit imposes
\begin{equation}
    \Delta {\cal R} = \Delta P^{(v)}_{z}(\tilde{\Gamma}).
\end{equation}
To evaluate $\Delta P^{(v)}_{z}(\tilde{\Gamma})$, we consider a continuous path $\mathscr{R}$ [see Fig.~\ref{fig:phasediagram} (d)], which is parameterized by 
\begin{eqnarray}
    (k_{x,y}, \phi) = (R_0\sin t, -R_0 \cos t),
\end{eqnarray}
with $R_0>0$ and $t\in[0,\pi]$. Apparently, $\mathscr{R}$ is gapped everywhere and the 1D polarization remains well-defined all along. For any $\beta>0$, $\mathscr{R}$ drives the evolution of $P^{(v)}_{z}(\tilde{\Gamma})$ from phase B to phase A, while avoiding the LSM singularity at $k_{x,y}=\phi=0$. By continuously tracking $P^{(v)}_{z}({\bf k}_\parallel)$ along $\mathscr{R}$ in Fig.~\ref{fig:phasediagram} (e), we find a quantized polarization jump $\Delta P^{(v)}_{z}(\tilde{\Gamma}) = -2$. This directly establishes a key conclusion of this work:
\begin{itemize}
    \item The QBT of LSM is a delicate TCP leading to $|\Delta{\cal R}|=2$.
\end{itemize}
As shown in Fig.~\ref{fig:phasediagram} (c), the transition between phases A and B can also be achieved by crossing the line of NLSM twice and passing through phase C. We thus expect the line of NLSM will support a $|\Delta{\cal R}|=1$ transition, which is numerically confirmed in the SM \ref{app:phaseC}. Note that a similar NLSM-induced RTP transition has also been reported in class-A systems~\cite{nelson:2021delicate}.

While the above discussions have focused on the LSM, all the conclusions in Sec \ref{sec:topoFS} and Sec \ref{sec:critical} also hold for the metal phase of the LK model with $|\lambda_1|>2|\lambda_2|$. This is because the identity term parameterized by $\lambda_1$ has no impact on the specific form of energy eigenstates and will further vanish at the QBT. Consequently, the delicate topology, which is a wavefunction property, is independent of the value of $\lambda_1$.

\section{Zoology of Delicate Topological Phases}
\label{sec:dti}

The $\Delta{\cal R}$ across the LSM has proved a delicate topological distinction for A and B. To clarify the explicit nature of each phase, however, we would require the topological band information over the entire BZ, which is beyond the scope of any continuum model. In this section, we will regularize the continuous Hamiltonian $H_{c}(\mathbf{k})$  on a 3D lattice to construct tight-binding models, with which a variety of delicate topological phases are uncovered. In particular, we will demonstrate the possibility for a stable topological insulator or semimetal to carry an additional built-in delicate topology and discuss the corresponding boundary physics.        

\begin{figure*}[t]
\centering
\includegraphics[width=2\columnwidth]{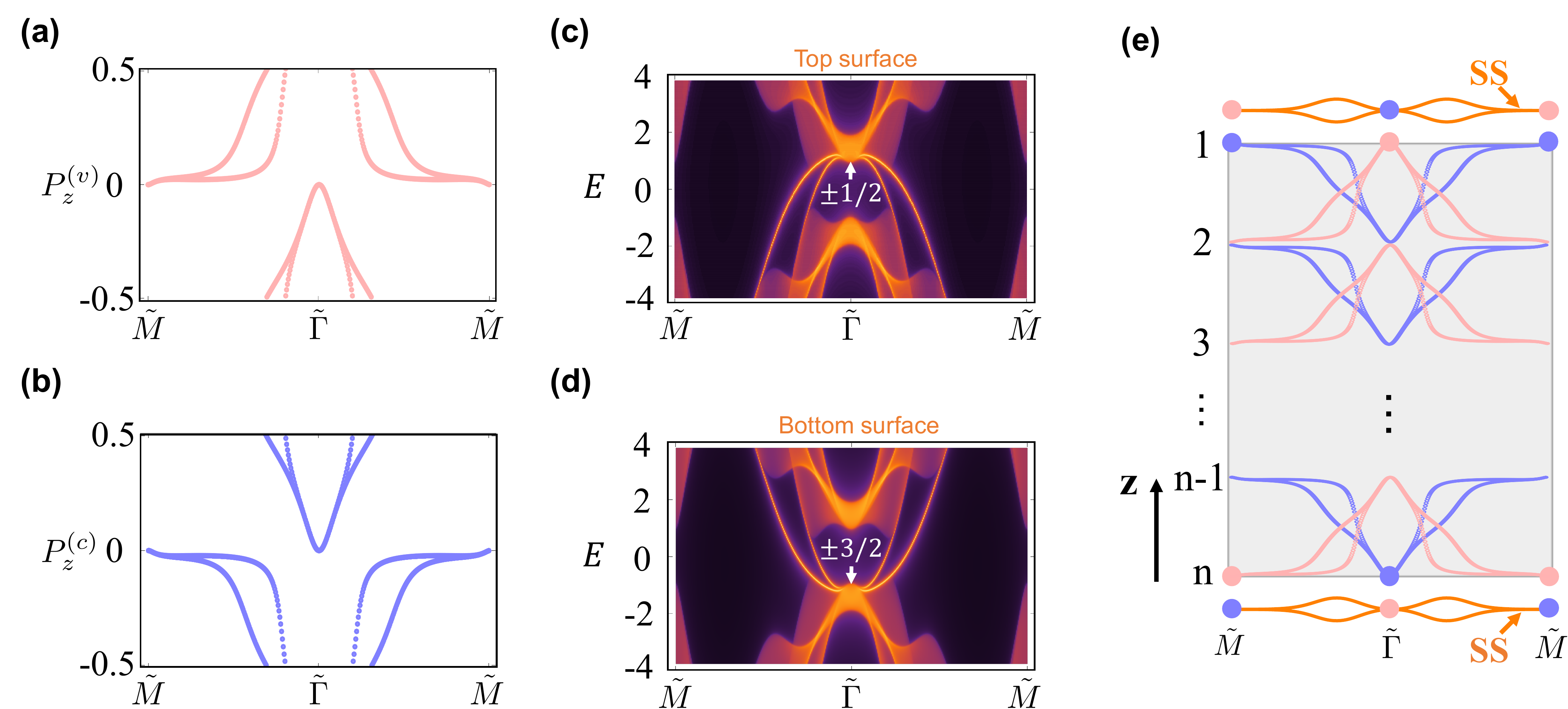}
\caption{The delicate TI phase of $\mathcal{H}_{a}(\mathbf{k})$ manifests in its Wilson loop spectra for valence (a) and conduction (b) electrons, where $|{\cal R}|=2$ is found. Energy spectra for the top and bottom (001) surfaces are shown in (c) and (d), respectively. Notably, the top and bottom surface states differ from each other in both their band connectivity and $z$-component angular momenta $J_z$ at $\tilde{\Gamma}$. We have marked the value of $J_z$ in the surface plots. These metallic surface modes are a direct consequence of the bulk-state RTP, as schematically explained in (e). Surface states with $J_z=\pm 3/2$ ($J_z=\pm 1/2$) are marked by red (blue) dots.
}
\label{fig:RTPinsulator}
\end{figure*}

\subsection{Delicate Topological Insulator}

We start by introducing a RTP-carrying delicate TI phase that is free from any stable topological physics. We will elaborate on how the bulk-state RTP will necessarily lead to anomalous surface states. To this end, we regularize $H_c({\bf k})$ into a lattice Hamiltonian ${\cal H}_a({\bf k})$ with the following rules,
\begin{eqnarray}
    && k_i \rightarrow \sin k_i,\ \ \ \ \ \forall i\in{x,y,z},  \nonumber \\
    && k_z^2 \rightarrow \sin^2 k_z,\ \ k_{x/y}^2 \rightarrow 2(1-\cos k_{x/y}).
    \label{eq:regularization_A}
\end{eqnarray}
Due to the lattice geometry, ${\cal H}_a({\bf k})$ is now invariant under a discrete point group of $C_{4v}$, as well as the time-reversal symmetry $\mathcal{T}$. Both symmetries are crucial for the delicate topology that we are going after. 

Notably, at $\Sigma=\phi=0$, ${\cal H}_a({\bf k})$ always exhibits a pair of four-fold degenerate QBTs at $\Gamma$ and $Z$, respectively. Each QBT resembles the one found in the continuum model $H_c({\bf k})$, and is thus expected to serve as a delicate TCP. For our purpose, we hope to spoil the simultaneous occurrence of these QBTs of ${\cal H}_a({\bf k})$, so that we can investigate their topological consequence individually. This motivates us to parameterize the Rashba parameter $\phi$ as
\begin{equation}
    \phi(\mathbf{k})=u-\sum_{i=x,y,z}\cos k_{i}.   \label{eq:phi_parameterization}
\end{equation}

Fixing $\Sigma=0$ and $\beta=1$, it is easy to see that $u=3$ leads to $\phi(\Gamma)=0$ and thus a QBT at $\Gamma$. For $2\sqrt{2}-1<u<3$, we find that ${\cal H}_a({\bf k})$ realizes a delicate TI with a RTP invariant of $\mathcal{R}=2$. Take $u=2$ as an example, the nontrivial RTP invariant for valence and conduction electrons are clearly visualized by tracking $P_{z}^{(v/c)}$ along $\tilde{M}-\tilde{\Gamma}-\tilde{M}$, as shown in Fig.~\ref{fig:RTPinsulator} (a) and (b), respectively. Moreover, it is easy to check that all valence (conduction) eigenstates of the lattice model have angular momenta $\pm 3/2$ ($\pm 1/2$) at both $\Gamma$ and $M$, thus satisfying the mutually disjoint condition. Note that this delicate TI with $\phi(\Gamma)<0$ corresponds to the phase B defined in Fig.~\ref{fig:phasediagram}. For $u>3$ and hence $\phi(\Gamma)>0$, we expect a phase A with ${\cal R}_A = {\cal R}_B-2 = 0$, which directly follows the quantized $P_z$ jump that the QBT can provide. Meanwhile, a gapped phase C with ${\cal R}=1$ can be achieved with a small negative $\Sigma$. While we focus on phase B here, discussions on phases A and C can be found in the SM \ref{app:phaseC}.  

The energy spectra for (001) top and bottom surfaces are calculated for the ${\cal R}=2$ phase using the iterative Green's function technique~\cite{Sancho_GreenF}, as shown in Fig.~\ref{fig:RTPinsulator} (c) and (d), respectively. On both surfaces, there exists a pair of non-Dirac surface states penetrating the bulk energy gap. However, the localized states on the opposite surfaces are highly asymmetric. For example, the top surface states feature a hole-like dispersion, with a $J_z=\pm 1/2$ Kramers pair sitting at the conduction band bottom at $\tilde{\Gamma}$. In contrast, the bottom ones are electron-like and they host a $J_z=\pm 3/2$ Kramers pair coinciding with the valence band top at $\tilde{\Gamma}$. The surface-state symmetry contents are obtained through matrix exact diagonalization in a thick slab geometry along the $z$ direction.   

This unusual boundary physics originates from the bulk-state RTP~\cite{nelson:2021delicate,nelson:2021multi,zhu:2023spinhopf}. Specifically, Fig.~\ref{fig:RTPinsulator} (a) implies that the hybrid Wannier centers of the valence electrons will be shifted by one unit cell along $+z$, as the system evolves from $\tilde{M}$ to $\tilde{\Gamma}$. The conduction electrons will behave in an exactly opposite way. This pumping process remains well-defined until it hits the surfaces. As shown in Fig.~\ref{fig:RTPinsulator} (e), the RTP in the valence subspace enforces the top layer ($z=1$) to feature a Kramers pair with $J_z=\pm 3/2$ at $\tilde{\Gamma}$, while that of the conduction electrons contribute to another Kramers pair with $J_z=\pm 1/2$ at $\tilde{M}$. However, the local Hilbert space at the top layer ($z=1$) always contains two Kramers pairs with $\pm 3/2$ and $\pm 1/2$ at both $\tilde{\Gamma}$ and $\tilde{M}$. As compensation, there must exist a Kramer pair of surface-localized states, which carries $J_z=\pm 1/2$ at $\tilde{\Gamma}$ and $J_z=\pm 3/2$ at $\tilde{M}$. Similar surface states with an opposite $J_z$ distribution will necessarily show up on the bottom surface as well. 

\begin{figure*}[t]
\centering
\includegraphics[width=2\columnwidth]{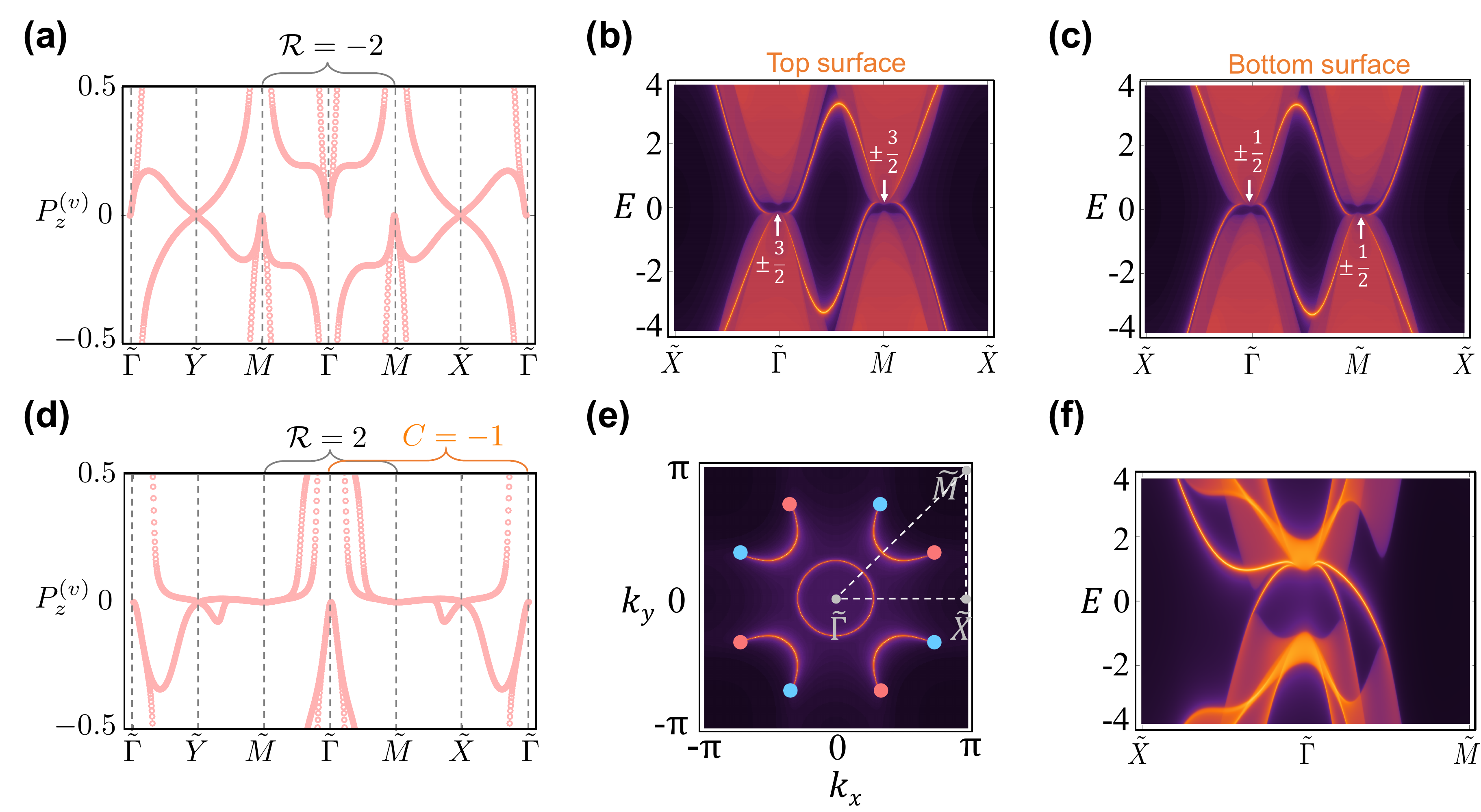}
\caption{(a) and (d) track the evolutions of $P_{z}^{(v)}$ in the RTP-enriched weak TI and WSM phases, respectively. The intertwined stable and delicate topologies are manifest in both cases. In (b) and (c), we show the (001) surface spectra for the weak TI phase, where we have marked the value of $J_z$ for the surface Dirac nodes. As for the WSM phase, we plot in (e) the Fermi surface for the top (001) surface. The red and blue dots are bulk Weyl nodes with a monopole charge of $+1$ and $-1$, respectively. (f) shows the energy spectrum for the top surface of the RTP-enriched WSM.}
\label{fig:RTPenrich}
\end{figure*}

Let us make several remarks on the surface states. First, they are guaranteed to be gapless when boundaries are ``ideally sharp"~\cite{nelson:2021multi}. This mechanism is intuitively addressed in Fig.~\ref{fig:RTPinsulator} (e), where the ``compensated" top surface states are mainly contributed by conduction electrons (blue dot) at $\tilde{\Gamma}$ and valence electrons (red dot) at $\tilde{M}$. The sharp surface limit implies that the energy of the surface state would be bounded by that of its bulk origin. As shown in Fig.~\ref{fig:RTPinsulator} (c), the top surface states will then be pinned to the valence-band top at $\tilde{M}$ and to the conduction-band bottom at $\tilde{\Gamma}$, necessarily penetrating the bulk energy gap. The energy dispersion for the bottom surface states can be understood similarly.   

Besides, the topological nature of these surface modes manifests in the symmetry contents of their wavefunctions. To illustrate, let us denote the effective 2D systems formed by the top/bottom surface states as $h_\text{SS}^{(t/b)}$. Clearly, $h_\text{SS}^{(t)}$ and $h_\text{SS}^{(b)}$ always feature an opposite $k$-space distribution of $J_z$, a phenomenon known as the angular-momentum anomaly. It is easy to tell from their band representations that $h_\text{SS}^{(t/b)}$ itself cannot be reproduced by any 2D local class-AII lattice model. Hence, the RTP-induced surface states are {\it anomalous}.

In a real-world material, there always exist many other bands beyond the delicate ones, so it is unlikely for the complete band manifold to respect the MDC. Therefore, delicate topology should, in principle, be interpreted as a low-energy effective phenomenon in a realistic system. However, as we numerically show in the SM~\ref{app:pertrubative}, the RTP physics is surprisingly robust against the multi-band effect. In particular, a visible breakdown of RTP occurs only when (i) certain types of band couplings are involved; and (ii) the coupling strength is sufficiently strong. Furthermore, the delicate topological surface states are more resilient against perturbations and persist even when the Wilson-loop spectrum of RTP is spoiled. This ``solid" nature of delicate topology hence makes the exotic surface physics predicted in this work much more accessible in experiments.

\subsection{RTP-Enriched Weak Topological Insulator}

With Eq.~\eqref{eq:regularization_A} in mind, we now consider a different and perhaps more natural regularization strategy to describe $H_c({\bf k})$ on a lattice, with
\begin{equation}
    k_i \rightarrow \sin k_i,\ \ k_i^2 \rightarrow 2(1-\cos k_i),\ \ \forall i=x,y,z.
\end{equation}
The new lattice model is termed ${\cal H}_b({\bf k})$, where we consider $\Sigma(\mathbf{k})=\Sigma_{0}\cos k_{z}$ and a constant Rashba parameter $\phi$. When $\phi=\Sigma_0=0$, ${\cal H}_b({\bf k})$ hosts a QBT at each of $\Gamma=(0,0,0)$ and $R=(\pi,\pi,\pi)$, respectively. For our purpose, we focus on the following choice of parameters $\beta=0$, $\phi=\sqrt{0.1}$, and $\Sigma_0=-0.1$. Fig.~\ref{fig:RTPenrich} (a) tracks the evolution of valence-band $P_z^{(v)}$ along the high-symmetry path in the rBZ, which clearly features a non-trivial RTP invariant $\mathcal{R}=-2$ along $\tilde{M}-\tilde{\Gamma}-\tilde{M}$.

Remarkably, a further examination of Fig.~\ref{fig:RTPenrich} (a) reveals the coexistence of $\mathbb{Z}_2$ topology. For example, the Wannier bands along $\tilde{Y} \rightarrow \tilde{\Gamma}$ feature exactly half of a helical winding pattern. Thanks to the time-reversal symmetry, the 2D plane defined by $k_x=0$ then carries a complete helical winding of $P_z^{(v)}$ and hence a $\mathbb{Z}_2$ index of $\nu_\text{2D}=1$~\cite{kane2005Z2}. Similarly, we find that $\nu_\text{2D}=1$ exists for every high-symmetry plane that is labeled by $k_{x,y,z}=0,\pi$. This immediately suggests that the lattice model ${\cal H}_b$ now carries a vanishing strong $\mathbb{Z}_2$ index and nontrivial weak indices~\cite{fu2007topo}, which is quantified as $(\nu_0; \nu_x, \nu_y, \nu_z) = (0; 1, 1, 1)$. Therefore, a pair of Dirac states will show up at $\tilde{\Gamma}$ and $\tilde{M}$ of the surface BZ for both top and bottom surfaces.

The interplay between RTP and $\mathbb{Z}_2$ physics has intriguing boundary consequences. Unlike the previous delicate TI phase, this weak TI phase satisfies the MDC independently at both $\tilde{\Gamma}$ and $\tilde{M}$. Specifically, all the valence states at $\tilde{\Gamma}$ have $J_z=\pm 3/2$, while their conduction counterparts feature $J_z=\pm 1/2$. The pattern of $J_z$ at $\tilde{M}$ is exactly the opposite of that at $\tilde{\Gamma}$. Following the bulk-boundary relation depicted in Fig.~\ref{fig:RTPinsulator} (e), the surface $J_z$ anomaly enforced by ${\cal R}=-2$ dictates that
\begin{enumerate}
    \item[(i)] The top (001) surface state at both $\tilde{\Gamma}$ and $\tilde{M}$ carries $|J_z|=\frac{3}{2}$, while the bottom one features $|J_z|=\frac{1}{2}$.
    \item[(ii)] The Kramers pair for the top surface is energetically pinned at the top of bulk valence bands at $\tilde{\Gamma}$, similar to Fig.~\ref{fig:RTPinsulator} (d), while that at $\tilde{M}$ sits at the bottom of bulk conduction bands. The bottom surface states behave exactly the opposite.  
\end{enumerate}
Since both delicate and $\mathbb{Z}_2$ topologies are shared by the same set of bulk bands, we thus expect the above exotic features will be carried by the Dirac surface states. This is directly confirmed in our iterative Green's function simulations in Fig.~\ref{fig:RTPenrich} (b) and (c). We have numerically plotted the energy spectra for both top and bottom (001) surfaces in a semi-infinite geometry, where a pair of Dirac states are found at $\tilde{\Gamma}$ and $\tilde{M}$, respectively. Remarkably, these surface phenomena exactly agree with our prediction from RTP.   

The weak TI state discussed above exemplifies the notion of RTP-enriched topological phases, where
\begin{enumerate}
    \item[(i)] Stable topology decides the number and $k$-space locations of the surface states.
    \item[(ii)] Delicate topology informs the energy and symmetry features of the surface states.
\end{enumerate}
This intriguing synergy between delicate and stable topologies arises from the fact that they are hosted by the same set of bands. On the other hand, it is also possible for delicate and stable topologies to coexist, but they are attributed to distinct band sets. This scenario falls out of the RTP-enriched topological physics defined above, and the delicate and stable topologies are thus expected to contribute independently to the boundaries. We leave the discussion of this kind of hybrid delicate-stable topological phases to future research.  

\subsection{RTP-Enriched Weyl Semimetal}

The notion of RTP-enriched topological physics also applies to gapless systems. As proof of concept, we will discuss a Weyl semimetal phase that emerges from the LK Hamiltonian. The lattice regularization scheme considered here is similar to Eq.~\eqref{eq:regularization_A}, with $k_i \rightarrow \sin k_i, \forall i\in{x,y,z}$. We dub this new lattice Hamiltonian ${\cal H}_c({\bf k})$, to distinguish it from the previous two lattice models. Note that the inversion symmetry is broken by the Rashba term $\phi(\mathbf{k})=u-\sum_{i=x,y,z}\cos k_{i}$, so Weyl fermions can, in principle, exist.  

To topologically characterize ${\cal H}_c({\bf k})$, we take $u=2$ and calculate $P_{z}^{(v)}$ along a high-symmetry $k$-path, as shown in Fig.~\ref{fig:RTPenrich} (d). It is evident the 2D plane $\tilde{M}-\tilde{\Gamma}-\tilde{M}$ hosts a RTP with ${\cal R}=2$. Furthermore, the torus defined by $k_{z}\times(\Gamma-M-X-\Gamma)$ clearly exhibits a quantized Berry flux characterized by a Chern number $C=-1$, which manifests in the chiral winding of Wannier bands. As a consequence, this torus must enclose a Weyl node of $+1$ monopole charge. Similarly, $\tilde{\Gamma}-\tilde{Y}-\tilde{M}-\tilde{\Gamma}$ should harbor a negatively charged Weyl node. Indeed, it is easy to check that the bulk dispersion of ${\cal H}_c({\bf k})$ features eight Weyl nodes living on the $k_z=0$ plane. Since both RTP and Weyl physics are hosted by the same set of bands, this represents a RTP-enriched Weyl semimetal phase. 

The value of ${\cal R}$ requires the emergence of two gapless surface states along $\tilde{\Gamma}-\tilde{M}$, with one of them contributed by the Weyl nodes. Consequently, there must exist another RTP-enforced surface state that is beyond the Weyl physics. To confirm our expectation, we exploit the iterative Green's function method to map out the zero-energy Fermi surface for (001) top surface, as shown in Fig.~\ref{fig:RTPenrich} (e). We can clearly observe the Fermi-arc surface states connecting Weyl points with opposite chiralities. Notably, the circular Fermi loop around $\tilde{\Gamma}$ is exactly the RTP-enforced surface state that we have predicted. We further plot the corresponding surface spectrum in Fig.~\ref{fig:RTPenrich} (f). We find that the Fermi-arc state features a saddle-point dispersion, which merges with the RTP surface state at $\tilde{\Gamma}$. Therefore, the role of RTP here is two-fold. It not only dictates the symmetry information and band connectivity of Fermi-arc states along $\tilde{\Gamma}-\tilde{M}$, but also enforces the existence of the extra surface state.

\section{Conclusions and Discussions}
\label{sec:conclusion}

To summarize, we have established the $j=3/2$ Luttinger systems as a realistic paradigm for delicate topology. Specifically, the QBT of LSMs not only carries a delicate topological charge, but also serves as a TCP connecting phases with distinct RTPs. Our lattice-model investigations clearly showcase the abundance of delicate topological phenomena that a LSM can lead to. We hope to specially highlight the new scenario of RTP-enriched topology, where the intertwinement between delicate and stable topologies will cause unusual boundary physics. This notion offers a new direction to explore the role of RTP in modern topological band theory.  

Material candidates of LSMs are common in nature, and notable examples include elemental $\alpha$-Sn~\cite{groves1963Tin}, pyrochlore iridates~\cite{kondo2015quadratic,cheng2017dielectric}, half-Heusler alloys~\cite{bansil2010heusler,yan2014half}, and many others~\cite{zhu2018quadratic,li2022LSM}. As RTP physics is generally incompatible with inversion symmetry, one can always exploit lattice strain or electric-field effects in the above candidates to induce the topological phases that we have predicted. On the other hand, existing non-centrosymmetric LSM systems may have naturally encoded RTP physics. For example, a LSM-based Weyl semimetal phase has been reported for Cu$_2$ZnGeSe$_4$~\cite{qian:2020weyl}, where we notice an unrecognized RTP hidden in the Wilson-loop spectrum. A similar mechanism for Weyl fermions also applies to strained HgTe~\cite{ruan2016symmetry} and half-Heusler alloys such as GdPtBi~\cite{hirschberger2016chiral,shekhar2018anomalous}. Further simulation efforts are needed to clarify the consequences of RTP in these Weyl candidates at the ab-initio level. 

As a final remark, real-world LSM materials often feature intriguing correlated electronic orders such as magnetism and superconductivity. The inherent delicate topology will shed new light on comprehending these interacting phenomena. For example, fragile topological bands are recently found to impose topological lower bounds for both superfluid weight~\cite{xie2020bound} and electron-phonon coupling strength~\cite{yu2023phonon}. It would be interesting to explore whether delicate topological bands in LSMs would feature similar effects. For superconducting LSMs, the delicate topological charge of $j=3/2$ QBT can, in principle, allow Cooper pairs to inherit a nontrivial two-body Berry phase and thus become nodal, which is similar to the scenarios discussed in WSMs~\cite{li:2018topological} and monopole NLSMs~\cite{yu:2022euler}. This mechanism could be highly relevant to half-Heusler compounds such as YPtBi~\cite{kim2018beyond}, where experimental evidence of nodal superconductivity is found. We leave a detailed study to future works~\cite{zhuzhangfuture}.  \\                     

{\it Acknowledgements} - We thank J. Yu, J. Sau, Y.-T. Hsu, Y.-X. Wang, Z. Bi, K. Sun, F. Schindler, R. Slager, and especially T. L. Hughes and A. Alexandradinata for stimulating discussions. RXZ is supported by a start-up fund of the University of Tennessee.  

\bibliography{apssamp}
% Produces the bibliography via BibTeX.

%%%%%%%%%%%%
\clearpage
\onecolumngrid
\setcounter{section}{0}
\setcounter{figure}{0}
\setcounter{equation}{0}
\renewcommand{\thefigure}{S\arabic{figure}}
\renewcommand{\theequation}{S\arabic{equation}}
\renewcommand{\thesection}{S\arabic{section}}
\begin{widetext}
\begin{center}
\textbf{\large Supplemental Material for ``Delicate Topology of Luttinger Semimetal"}
\end{center}

%%%%%%%%%%%%%%%%%%%%%%%%%%%%%%%%%%%%%%%%%%%%%%%%%%%%%%%%%%%%%%%
\section{Proof of Wannier Band Degeneracy in Eq.~\ref{eq:lambdaequator}}
\label{app:wlosymmetry}

The Wilson loop operator at latitude $\theta$ in valence/conduction space is a path-ordered product of projectors,
\begin{equation}
\hat{W}_{v/c}(k_{x0},k_{y0},\theta)=\prod_{\mathbf{k}\in \mathcal{L}_{\theta}} {\cal P}^{v /c}_{\mathbf{k}},
\end{equation}
where ${\cal P}^{v /c}_{\mathbf{k}}=\sum_{n\in v/c}\ket{u_{n}(\mathbf{k})}\bra{u_{n}(\mathbf{k})}$, and $(k_{x0},k_{y0})$ is the starting point of the path.
Let us consider a symmetry $g$ of the Bloch Hamiltonain such that $gH(\mathbf{k})g^{-1}=H(\mathscr{G}\mathbf{k}
)$, where $\mathscr{G}$ is the representation of $g$ in momentum space. This symmetry of the Hamiltonian will impose $gP^{v /c}_{\mathbf{k}}g^{-1}=P^{v /c}_{\mathscr{G}\mathbf{k}}$, further resulting in symmetry constraints of $\hat{W}_{v/c}(\theta)$.

It is important to note that for any $\alpha$, Hamiltonian $h^{(\alpha)}_{0}(\mathbf{k})$ [c.f. Eq.~\eqref{eq:generalizedLSM}] has an inversion symmetry $\mathcal{P}=\mathbbm{1}$, a time reversal symmetry $\mathcal{T}=i\sigma_{x}s_{y}K$, and a mirror-$z$ symmetry $m_{z}=\tau_{0}\sigma_{3}$ that anti-commutes with $\mathcal{T}$.
These symmetries will result in two constraints for conduction/valence Wilson loop operators along the equator, denoted as $\hat{W}_{c/v}^{\text{eq}}\equiv\hat{W}_{c/v}(\pi/2)$, which are useful to prove Eq.~\eqref{eq:lambdaequator} :
\begin{equation}
\label{eq:symmconstr}
\begin{aligned}
&\mathcal{P}\mathcal{T} \hat{W}^{\text{eq}}_{c/v}(\mathcal{P}\mathcal{T} )^{-1} =\sigma_{x}s_{y}(\hat{W}^{\text{eq}}_{c/v})^* \sigma_{x}s_{y}=\hat{W}^{\text{eq}}_{c/v},
\\
&m_{z}\hat{W}^{\text{eq}}_{c/v}m_{z}^{-1} =\hat{W}^{\text{eq}}_{c/v},
\end{aligned}
\end{equation}
The first constraint indicates that the unit-modulus eigenvalues of $\hat{W}^{\text{eq}}_{c/v}$ can only be $\pm 1$ or appear as a pair of complex conjugate numbers. For example, if $\ket{w_{v}}$ is an eigenstate of $\hat{W}^{\text{eq}}_{v}$ with eigenvalue $e^{i\lambda_{W,v}}$, then we have another eigenstate $\mathcal{P}\mathcal{T}\ket{w_{v}}$ with eigenvalue  $e^{-i\lambda_{W,v}}$:
\begin{equation}
\label{eq:con1}
\begin{aligned}
&\hat{W}^{\text{eq}}_{v}\ket{w_{v}}=e^{i\lambda_{W,v}}\ket{w_{v}},
\\
&\hat{W}^{\text{eq}}_{v}\mathcal{P}\mathcal{T}\ket{w_{v}}=\mathcal{P}\mathcal{T} \hat{W}^{\text{eq}}_{v}\ket{w_{v}}=e^{-i\lambda_{W,v}}\mathcal{P}\mathcal{T}\ket{w_{v}}.
\end{aligned}
\end{equation}
The second constraint indicates that each eigenstate of $\hat{W}^{\text{eq}}_{c/v}$ can simultaneously be an eigenstate of $m_{z}$ with an eigenvalue $m_{z}=+1$ or $-1$.
$\{\mathcal{P}\mathcal{T},m_{z}\}=0$ further indicates that $\ket{w_{v}}$ and $\mathcal{P}\mathcal{T}\ket{w_{v}}$ have opposite parities, i.e.,
\begin{equation}
\label{eq:con2}
\begin{aligned}
&m_{z}\ket{w_{v}}=\pm \ket{w_{v}}, 
\\
&m_{z}\mathcal{P}\mathcal{T}\ket{w_{v}}=-\mathcal{P}\mathcal{T}m_{z}\ket{w_{v}}=\mp\mathcal{P}\mathcal{T}\ket{w_{v}}.
\end{aligned}
\end{equation}
Hereafter, we label each eigenstate of a Wilson loop operator by the $M_z$ eigenvalue $m_{z}=\pm 1$, and denote the exponent of the eigenvalue in valence/conduction space as $\lambda_{W,c/v}^{\pm}$.
Then, from Eqs.~\eqref{eq:con1} and \eqref{eq:con2}, we can deduce
\begin{equation}
\label{eq:con3}
\lambda_{W,c/v}^{+}+\lambda_{W,c/v}^{-}=0 \ \text{modulo} \ 2\pi. 
\end{equation}
If we further consider the parallel transport over both conduction and valence bands of the $m_{z}=+1$ ($m_{z}=-1$) sector, then the Wilson loop operator is just an identity matrix with a determinant of $1$,
which indicates that
\begin{equation}
\label{eq:con4}
\begin{aligned}
\lambda_{W,v}^{+}+\lambda_{W,c}^{+}=0 \ \text{modulo} \ 2\pi,
\\
\lambda_{W,v}^{-}+\lambda_{W,c}^{-}=0 
\ \text{modulo} \ 2\pi.
\end{aligned}
\end{equation}
Combination of Eqs.~\ref{eq:con3} and \ref{eq:con4} immediately leads to the conclusion that the spectra of $\hat{W}_{v}$ and $\hat{W}_{c}$ are always bound together, i.e., $\lambda_{W,v}^{+}=\lambda_{W,c}^{-}=-\lambda_{W,v}^{-}=-\lambda_{W,c}^{+}$
modulo $2\pi$. This is the first part of Eq.~\eqref{eq:lambdaequator} in the main text.

To get an analytical expression for $\lambda_{W,c/v}^{\pm}$,we explicitly write down the $h_{0}^{(\alpha)}(k_{z}=0)$ under the basis where $m_{z}$ is diagonal:
\begin{eqnarray}
    h_0(\theta=0) = h_{+} \oplus h_{-},\ \ 
    h_\pm = \begin{pmatrix}
        \pm k_\parallel^2 & -\alpha \sqrt{3}k_-^2 \\
        -\alpha \sqrt{3} k_+^2 & \mp k_\parallel^2 \\
    \end{pmatrix},
\end{eqnarray}
where $k_{\parallel}\equiv \sqrt{k_{x}^2+k_{y}^2}$, and $h_{\pm}$ are blocks for $m_{z}=\pm 1$ sector, respectively. The valence eigenstate in $m_{z}=+1$ sector takes the form
\begin{equation}
\label{eq:eigenstate}
|\psi_v^{+} \rangle = \mathcal{N}(\frac{ 1- \sqrt{1+3\alpha^2}}{\sqrt{3}\alpha}e^{-2i\phi}, -1)^T, 
\end{equation}
where $\mathcal{N}=\frac{3\alpha^2}{2\sqrt{1+3\alpha^2}(\sqrt{1+3\alpha^2}-1)}$ is the normalization factor, and $e^{i\phi}\equiv k_{+}/k_\parallel$. From Eq.~\eqref{eq:eigenstate}, it is straightforward to calculate the Berry connection along the azimuthal direction:
\begin{equation}
\label{eq:azimuthalBC}
A_{\phi}=i\bra{\psi_{v}^{+}}\partial_{\phi}\ket{\psi_{v}^{+}}=1-\frac{1}{\sqrt{1+3\alpha^2}}.
\end{equation}
Since $\lambda^{+}_{W,v}$ is nothing but the Berry phase along the equator, then 
\begin{equation}
\label{eq:azimuthalBP}
\lambda^{+}_{W,v}=\int d\phi A_{\phi}=2\pi(1-\frac{1}{\sqrt{1+3\alpha^2}}).
\end{equation}
Eqs.~\eqref{eq:con4} and \eqref{eq:azimuthalBP} together give Eq.~\eqref{eq:lambdaequator} in the main text.

\section{$w_{\text{net}}$ as a delicate topological charge}
\label{app:delicacy}

We consider the $6$-band Kane model as a multi-band generalization of the LK Hamiltonian, which contains two extra $J_z= \pm 1/2$ $p$-orbital bands. Specifically,
\begin{equation}
\label{eq:kane}
H_{\text{Kane}}=\begin{pmatrix}
    h_{\frac{1}{2}} & T
    \\
    T^{\dag} & H_{LK}
\end{pmatrix},
\end{equation}
where $h_{\frac{1}{2}}=(E_{0} + \lambda_{3}k^2)\sigma_0$, and the coupling takes the form
\begin{equation}
\label{eq:T}
T({\bf k})=v\left(\begin{array}{cccc}
-\frac{1}{\sqrt{2}} k_{+} & \sqrt{\frac{2}{3}} k_z & \frac{1}{\sqrt{6}} k_{-} & 0 \\
0 & -\frac{1}{\sqrt{6}} k_{+} & \sqrt{\frac{2}{3}} k_z & \frac{1}{\sqrt{2}} k_{-}
\end{array}\right).
\end{equation}
Since $T(\Gamma)$=0, the QBT persists in the Kane model. We now numerically evaluate the Wilson loop spectrum on a Fermi sphere surrounding the QBT, of which the radius is $\pi/40$. In our calculations, we set $\lambda_{3}=-0.5$ and $E_{0}=-2$, and increase the coupling strength $v$ from $2$ to $6$. As shown in Fig.~\ref{fig:delicacy}, the Wilson-loop spectrum clearly violates Eq.~\eqref{eq:lambdaequator}, when the coupling between $h_{\frac{1}{2}}$ and $H_{LK}$ is sufficiently strong. This thus proves the ``delicacy" of $w_{\text{net}}$. 

\begin{figure}[h]
\centering
\includegraphics[width=0.9\columnwidth]{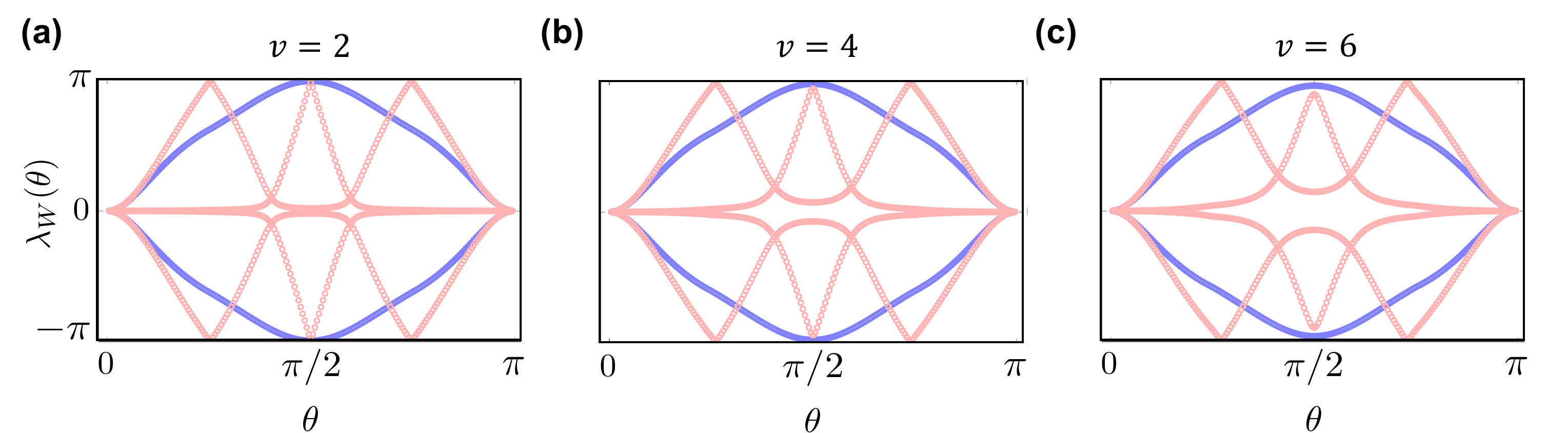}
\caption{The spectra of Wilson loop operators
versus $\theta$ in both valence and conduction spaces for the Kane model with (a) $v=2$; (b) $v=4$; and (c) $v=6$.}
\label{fig:delicacy}
\end{figure}

\section{Mutually disjoint condition and quantization of RTP}
\label{app:MDC}
In this section, we briefly review the relation between MDC and the quantization of RTP, following Refs.~\cite{nelson:2021multi,nelson:2021delicate,zhu:2023spinhopf,zhu:2023scattering}. We consider a tight-binding Hamiltonian written under a basis $\{\ket{\phi^{i}_{j}(\mathbf{R})}\}$, where $\mathbf{R}$ represents the unit cell coordinates, and  $i=1,2,\ldots$ labels different orbitals with angular-momentum $j$ in each unit cell. It is assumed that within one unit cell, basis states with same angular momentum share the same Wannier center. For clarity, the $z$-direction coordinate of the Wannier center for basis states with angular momentum $j$ is denoted as $w_{j}$. Additionally, we require that within one unit cell, all basis states, regardless of their angular momenta, have their Wannier center localized at the same rotation axis.

When the MDC is satisfied at a rotation invariant point $\tilde{K}_{0}$ in rBZ, we can draw two useful conclusions : (i) the $P_{z}^{(v)}$ at $\tilde{K}_{0}$ can be expressed as $P_{z}^{(v)}(\tilde{K}_{0})=\sum_{j\in \mathscr{J}_{v}(\tilde{K}_{0})}P_{z,j}^{(v)}(\tilde{K}_{0})$. Note that $\mathscr{J}_{v}(\tilde{K}_{0})$ is the set of rotation eigenvalues of all valence states at $\tilde{K}_{0}$, and $P_{z,j}^{(v)}(\tilde{K}_{0})$ is the polarization along $z$-direction contributed by all valence states at $\tilde{K}_{0}$ that have angular momentum $j$. (ii) $P_{z,j}^{(v)}(\tilde{K}_{0})\equiv N_{j} w_{j}+ n_{j}(\tilde{K}_{0})a$, where $N_{j}$ is the number of basis states with angular momentum $j$, $n_{j}(\tilde{K}_{0})$ is an integer, and $a$ is the lattice constant along $z$-direction. Point (i) is obviously true, so let us now explain point (ii). At a high symmetry point $\tilde{K}_{0}$ in the rBZ, the Bloch Hamiltonian is block-diagonal. Each of its  blocks, $H_{j}(k_{z},\tilde{K}_{0})$, can be viewed as a 1D tight-binding Hamiltonian in the symmetry sector labeled by $j$. The polarization of \emph{all} Bloch bands (i.e., both valence and conduction Bloch bands) of such a 1D tight-binding model is always the sum of the polarization of all basis states up to an integer multiple of lattice constant, i.e., $N_{j}w_{j}+n_{j}(\tilde{K}_{0})a$. This is because the Wannier orbitals constructed by inverse-Fourier transforming all Bloch bands are just the basis orbitals centered at $w_{j}$ to an integer number of lattice constants.  

Have established points (i) and (ii), it is clear to see that if MDC is satisfied at a pair of high symmetry points $\tilde{K}_{0}$ and $\tilde{K}_{0}^{\prime}$, then the polarization difference between them can be expressed as
\begin{equation}
P_{z}^{(v)}(\tilde{K}_{0})-P_{z}^{(v)}(\tilde{K}_{0}^{\prime})=a\left(\sum_{j\in \mathscr{J}_{v}(\tilde{K}_{0})}n_{j}(\tilde{K}_{0})-\sum_{j\in \mathscr{J}_{v}(\tilde{K}^{\prime}_{0})}n_{j}(\tilde{K}_{0}^{\prime})\right)+\left(\sum_{j\in \mathscr{J}_{v}(\tilde{K}_{0})}-\sum_{j\in \mathscr{J}_{v}(\tilde{K}_{0}^{\prime})}\right)N_{j}w_{j}.
\end{equation}
The above difference will be quantized as an integer in unit of lattice constant if the second term vanishes.  The second term will vanish if (i) $\mathscr{J}_{v}(\tilde{K}_{0})=\mathscr{J}_{v}(\tilde{K}^{\prime}_{0})$, or (ii) there are equal number of basis orbitals in each symmetry sector (i.e.,$N_{j}$ is independent of $j$), and all basis states come from the same orbital of the same atom in one unit cell (i.e.,$w_{j}$ is independent of $j$). Note that case (i) corresponds to the original MDC introduced in Ref.~\cite{nelson:2021delicate}, while case (ii) is naturally satisfied by the $j=3/2$ bands in perturbed LSM systems. Among the lattice models discussed in the main text, the delicate TI and WSM models satisfy both (i) and (ii), and the weak TI model only satisfies (ii). For simplicity, we uniformly refer to them as satisfying the MDC.

\section{Topological Phase Boundaries of $H_{c}({\bf k})$}
\label{app:gapless}

The energy spectrum of $H_{c}$ is given by
\begin{equation}
\label{eq: eHc}
E=\pm\sqrt{12(k_x^2+k_y^2)k_z^2+(k_x^2+k_y^2-2k_z^2+\Sigma-\beta\phi^2)^2+\left(\sqrt{12(k_x^2+k_y^2)\phi^2}\pm \sqrt{3}(k_x^2+k_y^2)\right)^2}.
\end{equation}
There is a gap closing if and only if the three terms under the square root are simultaneously zero. Let us start from the first term --$12(k_x^2+k_y^2)k_z^2=0$, which can be satisfied only when (i) $(k_x^2+k_y^2)=0$ and/or (ii) $k_{z}=0$. If (i) is true, then the third term automatically becomes zero, and the gap closing only requires the condition $-2k_{z}^2+\Sigma-\beta\phi^2=0$. This condition can be satisfied when $\Sigma\geqslant\beta\phi^2$, which corresponds to a DSM phase with two 4-fold-degenerate Dirac points at $k_{z}=\pm\sqrt{(\Sigma-\beta\phi^2)/2}$. If (i) is not satisfied but (ii) is satisfied, then the gap closing conditions are
\begin{equation}
\label{eq:condition}
k_x^2+k_y^2+\Sigma-\beta\phi^2=0, \ \sqrt{12(k_x^2+k_y^2)\phi^2}\pm \sqrt{3}(k_x^2+k_y^2)=0.
\end{equation}
It is straightforward to see that $\sqrt{12(k_x^2+k_y^2)\phi^2}+ \sqrt{3}(k_x^2+k_y^2)$ cannot be zero if (i) is not satisfied. Thus, we only need to consider the minus sign in the second equation. By solving Eq.~\eqref{eq:condition}, we derive a NLSM state with a 2-fold-degenerate nodal loop when $\Sigma=(\beta-4)\phi^2$. The nodal loop is in $k_{z}=0$ plane and is depicted by $k_x^2+k_y^2=4\phi^2$. Following the same process, one can also determine the gapless condition for lattice models that are lattice regularizations of $H_{c}$, and find those insulating and semimetallic topological phases discussed in Sec.~\ref{sec:dti} . 

\section{$\Delta\mathcal{R}$ between phase C and phases A/B}
\label{app:phaseC}
Here, we show that there is a jump in RTP, i.e., $|\Delta\mathcal{R}|=1$, between phase C and phases A/B in a lattice model. The lattice model we use corresponds to the lattice regularization of $H_{c}$ discussed in Eq.~\eqref{eq:regularization_A}, of which the Hamiltonian can be expressed as
\begin{equation}
\label{eq:RTPinsulator}
\begin{aligned}
H_{\text{RTPI}}({\bf k}) =& -2\sqrt{3} (\cos k_{y}-\cos k_{x}) \gamma_1 
- 2\sqrt{3} \sin k_x \sin k_y \gamma_2 -2\sqrt{3} \sin k_z \sin k_x \gamma_3 
\\
&- 2\sqrt{3} \sin k_z \sin k_y\gamma_4+M_{\text{RTPI}}(\mathbf{k})\gamma_5-2\sqrt{3}\phi(\mathbf{k}) (\sin k_x \gamma_{35} +\sin k_y \gamma_{34}),
\end{aligned}
\end{equation}
where $M_{\text{RTPI}}(\mathbf{k})=4-2\cos k_{x}-2\cos k_{y}-2\sin^2 k_{z}+\Sigma-\beta\phi(\mathbf{k})^2$, and $\phi(\mathbf{k})=u-\sum_{i=x,y,z}\cos k_{i}$. We then calculate $P_{z}^{(v)}$ over the $\tilde{M}-\tilde{\Gamma}-\tilde{M}$ in rBZ for $\beta=1$, and plot the results in Fig.~\ref{fig:phaseABC}. As we can observe, for $u=3.2$  ($u=2.8$) and $\Sigma=0$, which corresponds to phase A (B) shown in Fig~\ref{fig:phasediagram} (c), the lattice model has an RTP with value $\mathcal{R}=0$ ($\mathcal{R}=2$) along $\tilde{M}-\tilde{\Gamma}-\tilde{M}$ as shown in Fig.~\ref{fig:phaseABC} (a) (Fig.~\ref{fig:phaseABC} (b)). However,for $u=2.8$ and $\Sigma=-0.2<(\beta-4)[\phi(\Gamma)]^2=-0.12$, which corresponds to phase C shown in Fig~\ref{fig:phasediagram} (c), there is an RTP with value $\mathcal{R}=1$ as shown in Fig.~\ref{fig:phaseABC} (b). Thus, we can conclude that the jump of $\mathcal{R}$ upon the transition from phase C to phases A/B is $|\Delta \mathcal{R}|=1$. %It is also worth noting that in class-AII systems, the presence of RTPs with $\mathcal{R}=\pm 1$ directly signifies the existence of a nontrivial $\mathbb{Z}_2$ topology in the corresponding 2D plane [see Fig.~\ref{fig:phaseABC} (b) as an explicit example]. Thus, the transition between phases A/B and C can be understood as a hybrid transition of delicate topology and the stable, time-reversal protected $\mathbb{Z}_2$ topology. In contrast, the phase transition between phases A and B through the QBT of LSM is purely a topological transition of delicate topology.

\begin{figure}[h]
\centering
\includegraphics[width=0.9\columnwidth]{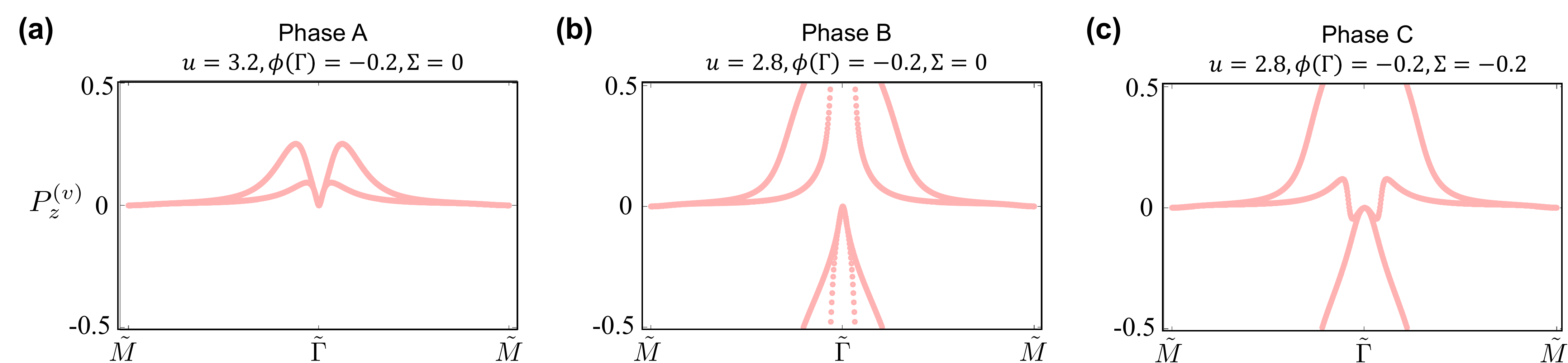}
\caption{(a), (b), and (c) show the RTP along $\tilde{M}-\tilde{\Gamma}-\tilde{M}$ for the lattice model in Eq.~\eqref{eq:RTPinsulator} with parameters in phase B, C, and A, respectively.}
\label{fig:phaseABC}
\end{figure}

\section{RTP-enriched weak topology: weak indices and RTP transition}
\label{app:weakTI}
\subsection{Weak $\mathbb{Z}_2$ indices}
Here, by parallel transporting the valence states along $k_{x}$ direction in $k_{z}=0$ and $k_{z}=\pi$ planes, we numerically calculate the $k_{y}$-dependent Wannier-bands in Fig.~\ref{fig:weakind}. Clearly, there are nontrivial $\mathbb{Z}_2$ indices in both planes, and thus the lattice model has a weak index of $1$ along $z$-direction. Together with the Wannier bands analysis in the main text, we can conclude that the lattice model is a weak $\mathbb{Z}_2$ TI with weak $\mathbb{Z}_2$ indices $(1,1,1)$.

\begin{figure}[h]
\centering
\includegraphics[width=0.7\columnwidth]{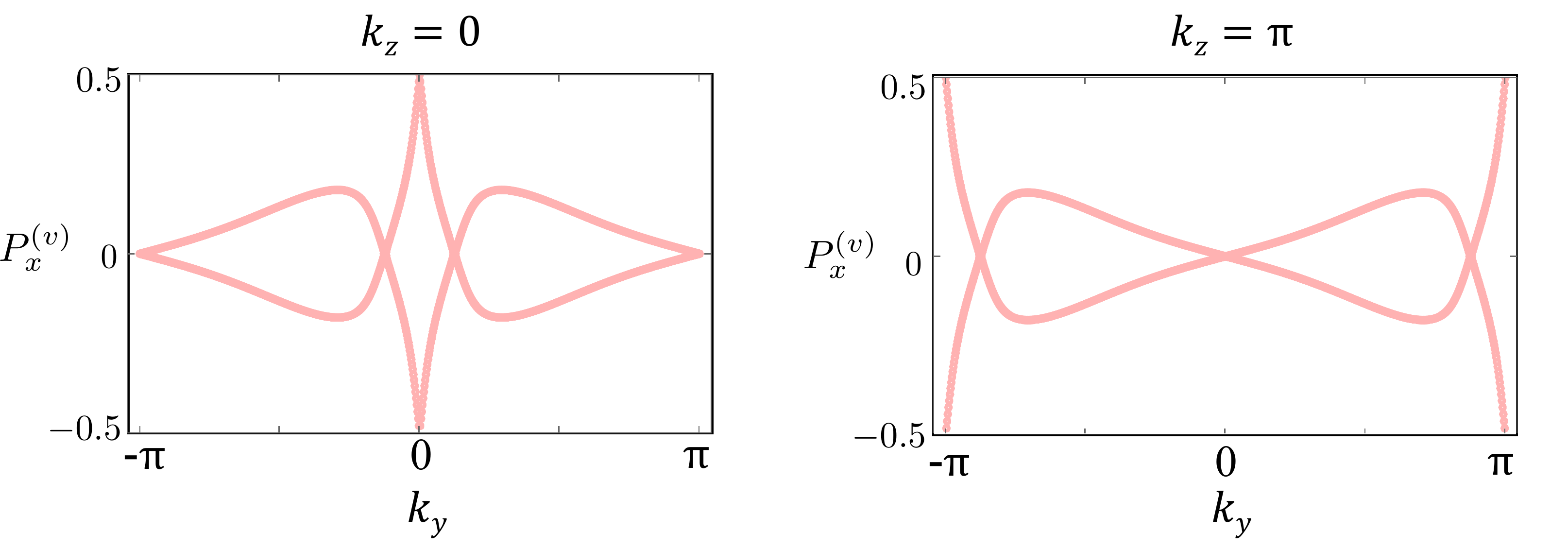}
\caption{Polarization along $x$-direction, $P_{x}^{(v)}$, versus $k_{y}$ in $k_{z}=0$ and $k_{z}=\pi$ planes.}
\label{fig:weakind}
\end{figure}

\subsection{Phase transition from $\phi>0$ to $\phi<0$}
In Fig.~\ref{fig:minusphi}, we plot the Wannier bands along the high symmetry lines, and the energy spectra showing the surface states on top and bottom surfaces for $\mathcal{H}_{b} (\mathbf{k})$ with $\phi=-\sqrt{0.1}$ and $\Sigma_{0}=-0.1$. Comparing to Fig.~\ref{fig:RTPenrich} (a), we can observe that from $\phi=\sqrt{0.1}$ to $\phi=-\sqrt{0.1}$, the RTP along $\tilde{M}-\tilde{\Gamma}-\tilde{M}$ jumps from $\mathcal{R}=-2$ to  $\mathcal{R}=2$, and thus $\Delta \mathcal{R}=4$. This is because at $\phi=0$ and $\Sigma_{0}=0$, the lattice model has two QBTs -- one at $\mathbf{k}=(0,0,0)$ and the other at $\mathbf{k}=(\pi,\pi,\pi)$, each of which contributes a $\Delta \mathcal{R}=2$. Since the RTP changes sign after the transition, the behaviors of surface states consequently exhibit a flip between the top and bottom surfaces, as shown in Fig.~\ref{fig:minusphi} (b) and (c).

\begin{figure}[h]
\centering
\includegraphics[width=0.9\columnwidth]{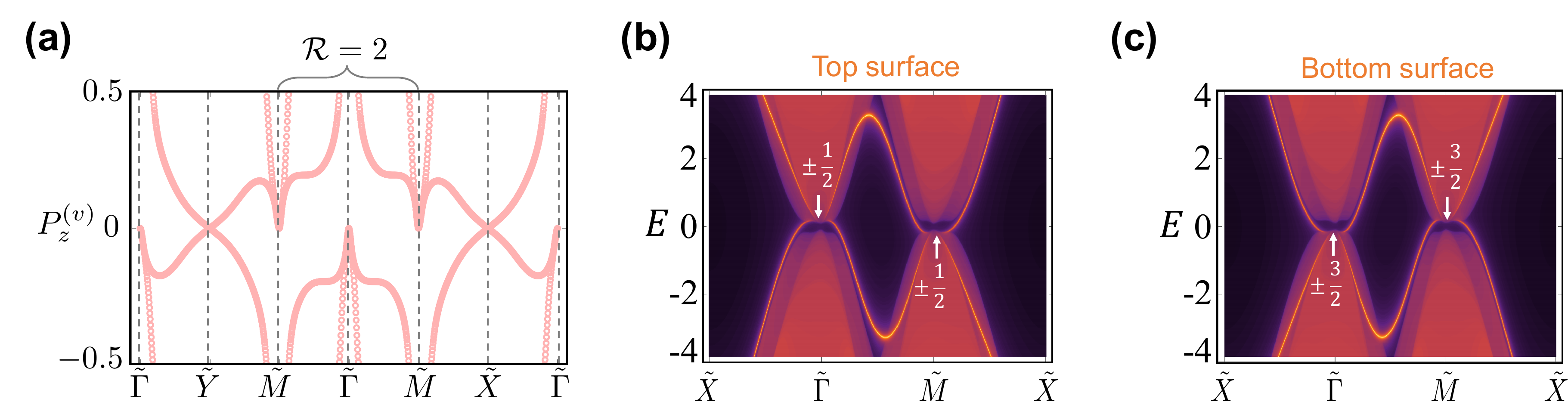}
\caption{For the weak TI model $\mathcal{H}_{b}$ with $\phi=-\sqrt{0.1}$ and $\Sigma_{0}=-0.1$, (a) shows $P_{z}^{(v)}$ along high symmetry lines. (b) and (c) respectively show the energy spectrum and the gapless surface states on the top and bottom surfaces.}
\label{fig:minusphi}
\end{figure}

\section{Robustness of delicate topology}
\label{app:pertrubative}

Here, we conduct a numerical study to understand how the inclusion of extra bands in a delicate topological insulator affects the RTP and the surface states. We begin by noting that adding $J_z=\pm 1/2$ bands following the Kane model presented in Eq.~\eqref{eq:kane} will not affect the RTP and the surface states at all. This is because the Kane model preserves an emergent 1D inversion symmetry along the rotation invariant axis, and thus the $P_{z}$ at $\tilde{M}$ and $\tilde{\Gamma}$ are forced to be quantized and cannot jump without bulk gap closing. A similar scenario has been previously discussed in Ref.~\cite{zhu:2023scattering}. 

To nullify the quantized RTP and illustrate the strength of this nullification under different conditions, we consider the addition of two flat bands in the valence space at energy $E_{\text{extra}}<0$ into the delicate TI model $\mathcal{H}_{a}(\mathbf{k})$. We choose the coupling between these two flat bands and the original bands as
\begin{equation}
\label{eq: T'}
T^{\prime}=t\left(\begin{array}{cccc}
0 &  e^{ik_{z}} & 0 & 0 \\
0 & 0 & e^{ik_{z}}& 0
\end{array}\right).
\end{equation}
which breaks the emergent 1D inversion symmetry. Explicitly, the 6-band Hamiltonian we study here takes the form
\begin{equation}
\label{eq:sixband}
H_{6}({\bf k})=\begin{pmatrix}
    h^{\prime}_{\frac{1}{2}} & T^{\prime}
    \\
    T^{\prime\dag} & \mathcal{H}_{a}(\mathbf{k})
\end{pmatrix},
\end{equation}
where $h^{\prime}_{\frac{1}{2}}=E_{\text{extra}}\sigma_{0}$.

We then tune the value of the two parameters $E_{\text{extra}}$ and $t$ to investigate the effects of the two extra bands.
As shown in Fig.~\ref{fig:extraband} (a), for the selected parameters $t=1$, corresponding to half of the bulk gap, and $E_{\text{extra}}=-1$, representing the energy at the valence band top, we clearly see that the RTP is no longer quantized. This is because the MDC is now violated, whereby the couplings between the two added $J_z=\pm 1/2$ bands and the pre-existing $J_z=\pm 1/2$ conduction bands spoil the quantization of $\mathcal{R}$. Nonetheless, in addition to their hybridization with the newly added bands, the surface states exhibit little alterations: they are still across the gap, and they still touch the conduction/valence bands at the high symmetry points as in the four-band case.  As we decrease the value of $t$ and $E_{\text{extra}}$ in Fig.~\ref{fig:extraband} (b) and (c), it is clear to observe that the effects from the two extra bands indeed become negligible, suggesting the robustness of RTP physics.

\begin{figure}[h]
\centering
\includegraphics[width=1\columnwidth]{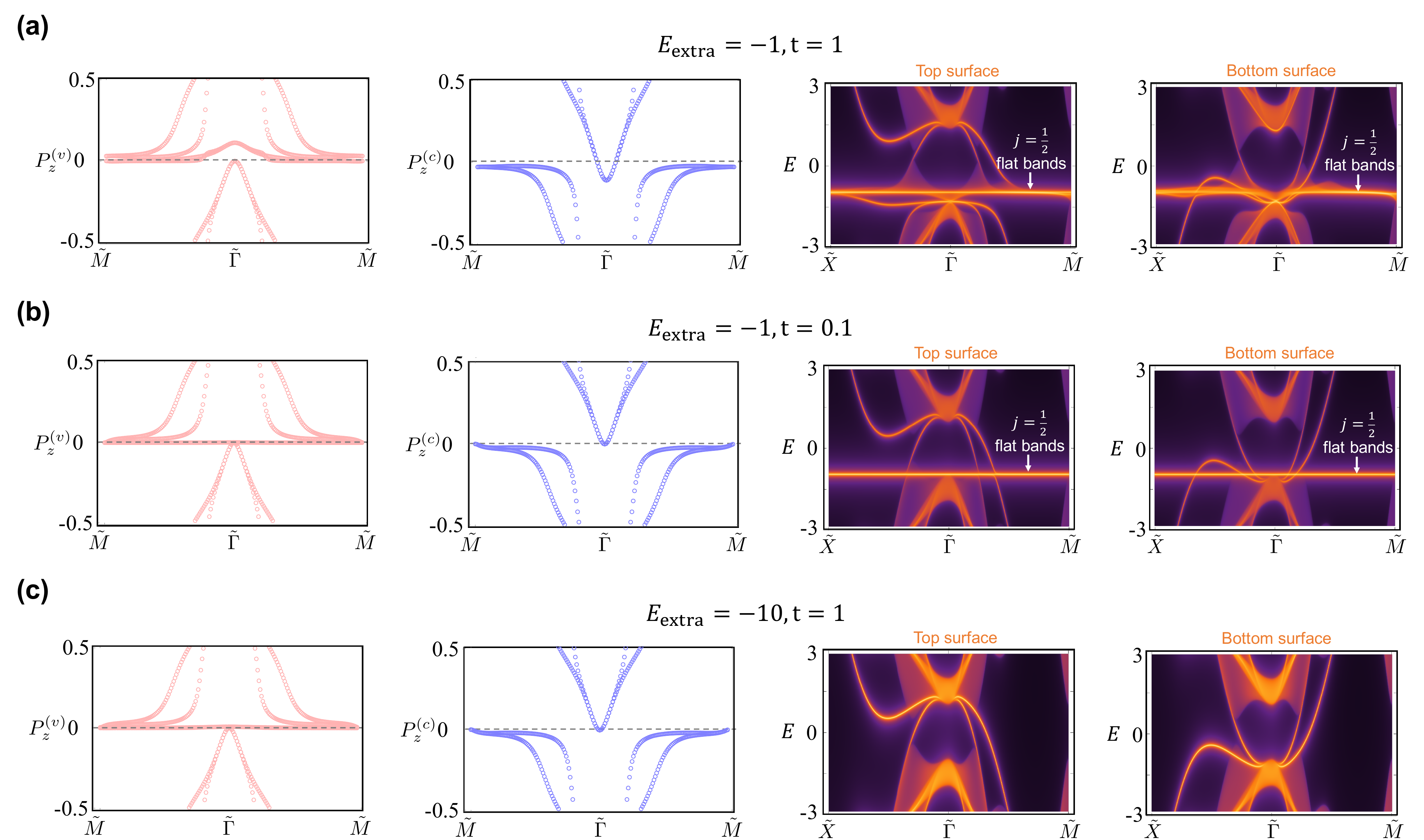}
\caption{RTP in valence/conduction spaces and surface states on top/bottom surfaces of $H_{6}$ with parameters (a) $E_{\text{extra}}=-1, t=1$; (b) $E_{\text{extra}}=-1, t=0.1$; and (c) $E_{\text{extra}}=-10, t=1$.}
\label{fig:extraband}
\end{figure}

\end{widetext}

\end{document}